\documentclass{appolb}
\usepackage{epsfig}

\usepackage{axodraw}
\usepackage{amsmath}
\usepackage{amssymb}
\usepackage{color}
\usepackage{graphics}
\usepackage{graphicx}
\usepackage{amsfonts}

\newcommand{\DM}{\Omega_{CDM}h^2}
\newcommand{\DMW}{\Omega^{WMAP}_{CDM}h^2}

\newcommand{\stau}{\tilde{\tau}}
\newcommand{\sel}{\tilde{e}_R}
\newcommand{\smu}{\tilde{\mu}_R}
\newcommand{\neut}{\tilde{\chi}^0_1}

\newcommand{\DeltaO}{\Delta^{\Omega}}

\newcommand{\tanb}{\tan\beta}

\begin{document}
\title{Natural Dark Matter \thanks{Presented at Physics at the LHC
    2006, Cracow.}  } \author{S.~F.~King,~J.~P.~Roberts
  \address{School of Physics and Astronomy,
    University of Southampton,\\
    Southampton, SO17 1BJ, U.K.}  } \maketitle
\begin{abstract}
  In this talk we analyse the claim that supersymmetry (SUSY)
  naturally accounts for the observed dark matter density. In many
  cases, it is necessary to tune the parameters of a SUSY model to fit
  the WMAP data. We provide a quantitative analysis of the degree of
  tuning required for different annihilation channels. Some regions
  are natural, requiring no tuning at all, whereas others require
  tuning at the $0.1\%$ level.
\end{abstract}
  
\section{Introduction}
A key motivation for TeV scale supersymmetry (SUSY) is that it
provides a natural dark matter candidate if the lightest neutralino is
the LSP. However the regions of parameter space that yield neutralino
dark matter in agreement with WMAP look very restricted. We recently
studied the naturalness of dark matter in \cite{hep-ph/0603095}.

Questions of fine-tuning have long been considered in the case of
electroweak symmetry breaking. In many of these studies the degree of
fine-tuning required was quantified through a measure of the
sensitivity of $m^2_Z$ to the input parameters of the MSSM
$a_{MSSM}$. We use a similar measure \cite{ellis} to quantify the degree
of fine-tuning required of the MSSM parameters to produce an LSP that
reproduces the observed dark matter relic density:
\begin{equation}
  \Delta_{a_{MSSM}}^{\Omega}=\left|\frac{\partial
    \ln\left(\DM\right)}{\partial \ln\left(a_{MSSM} \right)}\right|.
  \label{meas}
\end{equation}
We take the total tuning of a point to be
$\DeltaO=\text{max}\left(\DeltaO_a\right)$.

The calculation of $\DM$ primarily depends on $a_{MSSM}$ through their
effect on the annihilation cross-section of the lightest neutralino
$\neut$. This is primarily determined by the mass and composition of
$\neut$. This in turn is determined by diagonalising the neutralino
mass matrix at low energy. The matrix depends upon $M_1$, $M_2$, and
$\mu$. If one of these is much lighter than the others $\neut$ will be
primarily of that form. This allows us to divide up the MSSM parameter
space depending on the composition of $\neut$. A wino LSP ($M_2\ll
M_1,~\mu$) or higgsino LSP ($\mu\ll M_1,~M_2$) annihilates very
efficiently, resulting in too little dark matter, $\DM\ll \DMW$. A
bino LSP ($M_2\ll M_1,~\mu$) annihilates primarily via t-channel
slepton exchange. This process is only efficient for light sleptons
and so bino LSPs generally result in too large a relic $\DM \gg \DMW$.

Therefore to fit the observed dark matter density we are required to
move to unusual regions of the parameter space. One possibility is to
consider a mixed LSP. If we have a bino LSP with just enough of either
wino or higgsino mixed in, we can fit $\DMW$. This is the so called
``well-tempered'' neutralino championed in \cite{hep-ph/0601041}.
Alternatively we can consider a bino LSP in which the annihilation
cross-section is enhanced via some means. This can occur in a few
different ways. Firstly, if there are light sfermions, t-channel
sfermion is enhanced. Secondly if $2m_{\neut}=m_{Z,A,h}$ the
neutralinos can annihilate to an on-shell boson. Finally if the NLSP
is quasi-degenerate in mass with the LSP, there will be a significant
NLSP number density at freeze out and we must factor in annihilations
of the NLSP into our calculations of the SUSY relic density. All of
these effects can enhance annihilation of a bino LSP to the extent
that we fit the observed dark matter relic density. We would expect
each region to exhibit a different sensitivity to the MSSM input
parameters. To study these regions we take 4 different sets of
boundary conditions on the MSSM input parameters $a_{MSSM}$ at
$m_{GUT}$, beginning with the familiar case of the constrained minimal
supersymmetric standard model (CMSSM).

The rest of this talk is set out as follows. In section \ref{CMSSM}
we consider the CMSSM, which provides us with a useful reference point
against which the subsequent non-universal cases may be compared. In
section \ref{Scalar} we allow the third family soft sfermion and Higgs
mass squared to vary independently. In section \ref{Gauge} we consider
neutralino dark matter with non-universal gaugino masses, but with a
universal soft scalar mass.  In section \ref{ScalarGauge} we consider
{\em both} the effects of including an independent third family
sfermion mass squared {\em and} non-universal soft gaugino masses.
Section \ref{Conc} concludes the talk.

\section{CMSSM}
\label{CMSSM}

The CMSSM has 4 free inputs:
\begin{equation}
a_{CMSSM}\in\left\{m_0,~m_{1/2},~A_0,~\tanb\text{ and sign}(\mu)\right\}.
\end{equation}
$m_0$ is a common scalar mass that sets the soft masses of the
sfermion and Higgs sectors. $m_{1/2}$ is a common gaugino mass. $A_0$
sets the soft SUSY breaking trilinear coupling. $\tanb$ is the ratio
of the Higgs VeVs. Finally the requirement that the model provide
radiative electroweak symmetry breaking determines the magnitude of
the SUSY conserving Higgs mass $\mu$ but leaves the sign as a free
parameter.

The mass and composition of the lightest neutralino is determined by
diagonalising a mass matrix that depends upon the parameters
$M_1,~M_2\text{ and }\mu$ at the electroweak scale. In the CMSSM
$M_1=M_2=m_{1/2}$ at the GUT scale. As running effects mean that
$M_1(m_Z)\approx 0.4M_1(m_{GUT})$ and $M_2(m_Z)\approx
0.8M_2(m_{GUT})$, gaugino mass unification sets $M_1(m_Z)\approx 0.5
M_2(m_Z)$. Therefore, unless $\mu<M_1$, the lightest neutralino will
be dominantly bino.
\begin{figure}[t]
  \begin{center}
    \scalebox{.4}{\includegraphics{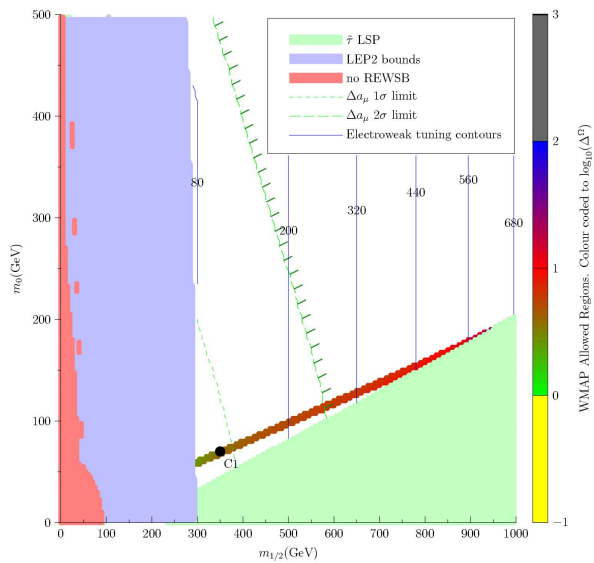}}
  \end{center}
  \caption{\footnotesize{The $(m_{1/2},m_0)$
    plane for the CMSSM with $A_0=0$,
    $\tan\beta=10$.}\label{fig:CMSSM}}
\end{figure}

In Fig.~\ref{fig:CMSSM} we consider the $(m_0,m_{1/2})$ plane of the
CMSSM parameter space with $\tanb=10,~A_0=0$. Across this parameter
space $\neut$ is bino. This generally results in $\DM\gg\DMW$.
However at low $m_0$ the $\stau$ becomes light. In the light (green)
region $m_{\stau}<m_{\neut}$ and the region is ruled out as this would
result in a charged LSP. Along the edge of this region
$m_{\stau}\approx m_{\neut}$. This means that at the time of freeze
out there would have been a large number density of $\stau$ alongside
the $\neut$, allowing many more annihilation channels than are open
for neutralinos alone. This results in a significant decrease of the
neutralino relic density. In the multicoloured strip that lies
alongside the $\stau$ LSP region, this coannihilation process results
in $\DM=\DMW$. The varying colours of this strip represent the value
of $\DeltaO$, defined by the colour legend on the right.

Coannihilation occurs when the LSP and NLSP are close in mass. As a
result, the efficiency of coannihilation processes is highly sensitive
to the mass difference $\delta m=m_{NLSP}-m_{LSP}$. If these masses
are determined by separate parameters we would expect that a large
degree of tuning would be required to fit the observed dark matter
density. In the coannihilation strip of Fig.~\ref{fig:CMSSM} the NLSP
is the stau. The stau mass is set at the GUT scale by $m_0$ and the
neutralino mass is set by $m_{1/2}$. As these are independent
parameters, we would expect the $\stau-\neut$ coannihilation region to
exhibit considerable fine-tuning.

The colour coding of the coannihilation strip shows a tuning of
$3-15$, considerably lower than would be expected if $m_{\stau}$ and
$m_{\neut}$ were unrelated. The smallness of the tuning comes from the
fact that along the coannihilation strip $m_0<m_{1/2}$. The running of
the right handed slepton masses are strongly dependent on $M_1$. When
$m_0$ is small, the dominant contribution to the low energy $\stau$
mass is via this running contribution from $M_1$. Thus in this region
of the CMSSM $m_{\stau}$ depends strongly on $m_{1/2}$, resulting in a
correlation of the masses of the neutralino and the stau at low
energy. It is this correlation of the masses that results in the low
tuning observed. 

Though we do not show it here, we have also investigated the other
regions of the CMSSM that fit $\DMW$. For large $m_0$ $\mu$ becomes
small and we have a bino/higgsino LSP. We find that such regions
exhibit a tuning $\DeltaO= 30-60$, less natural than the
coannihilation strip. For large $\tanb$ we can also access a region in
which $m_{A}\approx 2m_{\neut}$. This allows for neutralino
annihilation via the production of an on-shell pseudoscalar Higgs
boson. We find such an annihilation channel to require a tuning
$\DeltaO=80-300$. Finally at large $\tanb$ the running of the $\stau$
mass is no longer dominated by $M_1$. This breaks the correlation
between $m_{\neut}$ and $m_{\stau}$ at low energies and results in the
coannihilation strip that requires a tuning $\DeltaO\approx 50$.  This
bears out our expectations that coannihilation should require
significant tuning to achieve in normal circumstances.

\section{Non-universal Scalars}
\label{Scalar}

Our first move away from the universality of the CMSSM is to relax the
universality between the generations of sfermions, setting the soft
masses to be:
\begin{eqnarray*}
  m^2_{\tilde{Q}},m^2_{\tilde{L}},m^2_{\tilde{u}},m^2_{\tilde{d}},m^2_{\tilde{e}}&=&
  \begin{pmatrix}
    m^2_0 & 0 & 0 \\ 0 & m^2_0 & 0 \\ 0 & 0 & m^2_{0,3}
  \end{pmatrix}, \\
  m^2_{H_u} = m^2_{H_d} &=& m_{0,3}^2, \\ M_\alpha &=& m_{1/2}.
\end{eqnarray*}

This allows us to vary the 3rd family sfermion and Higgs mass squareds
separately from the 1st and 2nd families. This allows us to have light
1st and 2nd family sfermions without violating LEP bounds on the
lightest Higgs. In such a region we will also have a normal mass
hierarchy (NMH) in which the 1st family sfermions are the lightest and
the 3rd family sfermions are the heaviest, in contrast to the inverted
mass hierarchy found in the case of universal soft scalar masses. 
\begin{figure}[t]
  \begin{center}
    \scalebox{.4}{\includegraphics{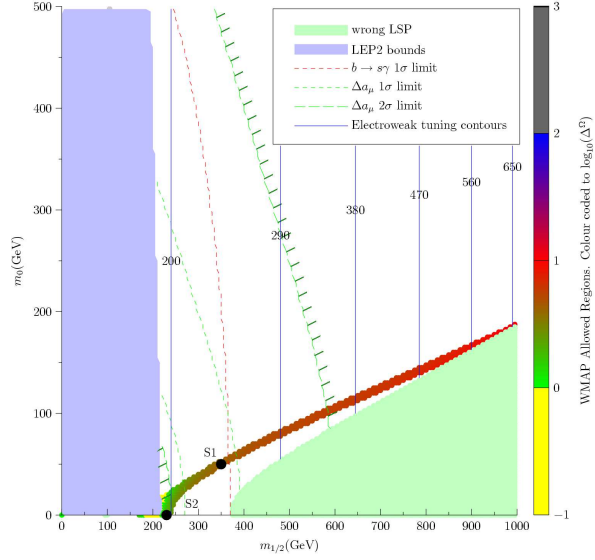}}
  \end{center}
  \caption{\footnotesize{The $(m_{1/2},m_0)$ plane for non-universal sfermion
    masses with $m_{0,3}=1$ TeV, $A_0=0$,
    $\tan\beta=10$.} \label{fig:Scalar1}}
\end{figure}

In Fig.~\ref{fig:Scalar1} we display the $(m_0,m_{1/2})$ plane for
$m_{0,3}=1$ TeV, $A_0=0$, $\tan\beta=10$. By increasing the soft mass
of the Higgs bosons to $1~\text{TeV}$ the LEP bound has moved down to
$200~\text{GeV}$. As before we have a coannihilation strip but as
$m_{\stau}>m_{\sel,\smu}$, the coannihilation here is with selectrons
and smuons rather than the stau.

As before the coannihilation strip exhibits a tuning $\DeltaO\approx
10$ across much of its length. This drops to $\DeltaO\approx 2$ for
$m_{1/2}<260~\text{GeV}$. This decrease has two causes. Firstly, for
$m_{1/2}<370~\text{GeV}$ we can access $m_0=0$.  For $m_0\approx 0$,
the mass of $\sel$ and $\smu$ are almost entirely determined by the
running effects from $M_1$ resulting in a strong correlation between
$m_{\sel,\smu}$ and $m_{\neut}$. This decreases the tuning required to
provide coannihilation. Secondly, as we move to low $m_0$ and
$m_{1/2}$, we decrease the mass of the sleptons themselves, enhancing
neutralino annihilation via t-channel slepton exchange. Indeed at
point S2 t-channel slepton exchange accounts for $60\%$ of the
annihilation. The cross-section for t-channel slepton exchange varies
slowly with the mass of the exchanged slepton and is relatively
insensitive to other parameters. Therefore it requires little or no
tuning to achieve. By maximising annihilation via t-channel slepton
exchange we minimise the required tuning.

\section{Non-universal Gauginos}
\label{Gauge}

We now relax the constraint of universal gaugino masses. By allowing
$M_1$, $M_2$ and $M_3$ to vary independently we can control the bino/wino
mixture of $\neut$. $M_3$ also has a strong effect on the running of
the Higgs masses so by keeping $M_3$ large we can avoid the LEP bound
on the lightest Higgs.
\begin{figure}[t]
  \begin{center}
    \scalebox{.4}{\includegraphics{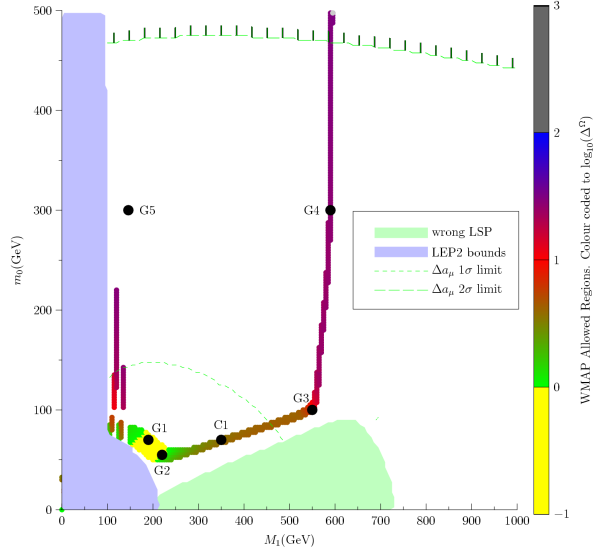}}
  \end{center}
  \caption{\footnotesize{The $(M_1,m_0)$ plane for non-universal gauginos
    with $M_2=M_3=350$ GeV, $A_0=0$,
    $\tan\beta=10$.}\label{fig:Gauge1}}
\end{figure}

In Fig.~\ref{fig:Gauge1} we show the $(m_0,M_1)$ plane for
$M_2=M_3=350~\text{GeV}$, $A_0=0$ and $\tan\beta=10$. In contrast to
previous figures there are a large number of strips that agree with
$\DMW$. The strip running from G2 to G3 is the stau coannihilation
strip we have seen before. The tuning of this strip agrees with our
previous findings in the case of the CMSSM. The strip that runs from
G3 through G4 corresponds to a well-tempered bino/wino neutralino.
As this strip exhibits a tuning of order 30 we once again
find that such ``well-tempered'' regions are less natural than
coannihilation strips.  At low $M_1$ there are two broken vertical
lines. These correspond to neutralino annihilation via the production
of an on-shell $Z$ or $h^0$. The $h^0$ resonance stretches to
$m_0>500~\text{GeV}$ but is too thin for this plot to resolve. This
channel requires tunings $\DeltaO=10-1000$ and so cannot be considered
natural. Finally, the yellow region that incorporates the point G1
represents annihilation solely via t-channel slepton exchange. This
region requires a tuning $\DeltaO<1$ and represents supernatural dark
matter.

\section{Non-universal Scalars and Gauginos}
\label{ScalarGauge}

Finally we relax both the universality of the gaugino masses and the
universality between sfermion generations at the same time. This
allows us to test the robustness of our findings in each case against
further non-universality. It also allows us to study a region in
which $M_2(m_Z)\approx \mu$ and the lightest neutralino is a
``maximally-tempered'' bino/wino/higgsino.
\begin{figure}[t]
  \begin{center}
    \scalebox{.4}{\includegraphics{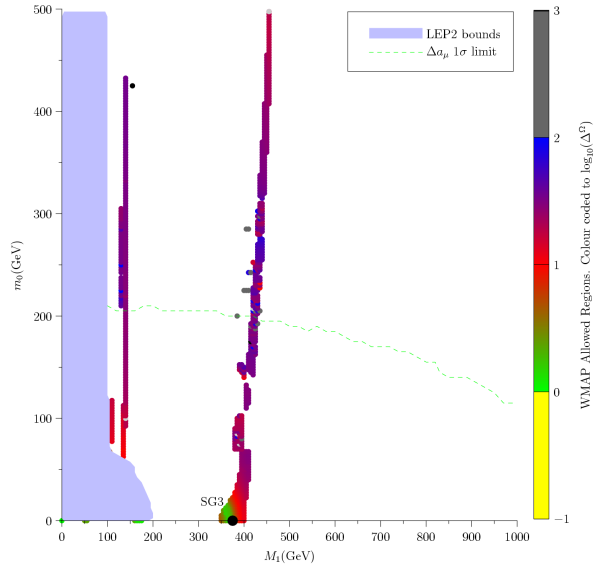}}
  \end{center}
  \caption{\footnotesize{The $(M_1,m_0)$ plane for non-universal gaugino and
    sfermion masses with $M_2=M_3=350\ \text{GeV}$,
    $m_{0,3}=2250\ \text{GeV}$, $A_0=0$,
    $\tan\beta=10$.}\label{fig:ScalarGauge2}}
\end{figure}

In Fig.~\ref{fig:ScalarGauge2} we show the $(m_0,M_1)$ plane with
$M_2=M_3=350~\text{GeV}$, $m_{0,3}=2250~\text{GeV}$, $A_0=0$ and
$\tan\beta=10$. At low $M_1$ we find the $h^0$ and $Z$ resonances as
before. The new feature is the line at $M_1=400~\text{GeV}$ that
incorporates the point SG3. Throughout this region $\neut$ is a mix of
bino, wino and higgsino. At point SG3 the wino and higgsino components
are roughly equal resulting in a maximally-tempered neutralino. This
region exhibits a low tuning $\DeltaO=4$, considerably below that
required for either bino/wino or bino/higgsino neutralinos.

\section{Conclusions}
\label{Conc}

We have studied the naturalness of the annihilation channels that
allow MSSM neutralinos to account for the observed dark matter
density. Within the four different sets of GUT scale boundary
conditions considered, these annihilation channels each display
characteristic degrees of fine-tuning. The largest tunings ($\DeltaO$
up to $1000$) appear for annihilation via on-shell production of Higgs
bosons. Moderate tunings $\mathcal{O}(30-60)$ are required for
``well-tempered'' neutralinos or slepton coannihilation with
uncorrelated masses. The most natural annihilation channel is
annihilation via t-channel slepton exchange. We have also found that
certain RGE effects can result in surprising drops in the tuning for
different channels. This is clearest in the case of slepton
coannihilation for low $m_0$ and $\tanb$. In this case the mass of
both particles is dominated by $M_1$ and the low energy masses are
correlated. This results in almost natural coannihilation, which
refutes the conclusions of \cite{hep-ph/0601041}.

These results have recently been extended to the case of a type I
string inspired model \cite{hep-ph/0608135}. In such a model the input
parameters differ from those of the MSSM and the characteristic
tunings of different annihilation channels can vary.

\end{document}